\documentclass[prl,reprint,showpacs]{revtex4-1}

\usepackage{amssymb,amsfonts,amsmath} 
\usepackage{graphicx}
\usepackage{color}
\usepackage{hyperref}

\begin{document} \title{Oscillatory dynamics and non-markovian memory in dissipative quantum systems}

\author{D.M.\ Kennes}  
\affiliation{Institut f{\"u}r Theorie der Statistischen Physik, RWTH Aachen University 
and JARA---Fundamentals of Future Information
Technology, 52056 Aachen, Germany}

\author{O.\ Kashuba}  
\affiliation{Institut f{\"u}r Theorie der Statistischen Physik, RWTH Aachen University and JARA---Fundamentals 
of Future Information
Technology, 52056 Aachen, Germany}

\author{M.\ Pletyukhov}  
\affiliation{Institut f{\"u}r Theorie der Statistischen Physik, RWTH Aachen University and JARA---Fundamentals 
of Future Information
Technology, 52056 Aachen, Germany}

\author{H.\ Schoeller}  
\affiliation{Institut f{\"u}r Theorie der Statistischen Physik, RWTH Aachen University and JARA---Fundamentals 
of Future Information
Technology, 52056 Aachen, Germany}

\author{V.\ Meden} 
\affiliation{Institut f{\"u}r Theorie der Statistischen Physik, RWTH Aachen University and JARA---Fundamentals 
of Future Information
Technology, 52056 Aachen, Germany}

\begin{abstract} 
The nonequilibrium dynamics of a small quantum system
coupled to a dissipative environment is studied. We show that
(1) the oscillatory dynamics close to a coherent-to-incoherent
transition is significantly different from the one of the 
classical damped harmonic oscillator and that (2) non-markovian 
memory plays a prominent role in the time evolution after
a quantum quench.

\end{abstract}

\date{\today} 
\maketitle

To describe dissipation in open quantum systems one often relies on 
phenomenological approaches. This might be sufficient to model 
experimental observations as long as the system is large, the coupling to the environment 
poorly specified, and the accuracy of the measurement limited. The rapid 
progress in controlled access to small quantum systems in such 
distinct fields as condensed matter physics, quantum optics, 
physical chemistry, and quantum information science renders a more microscopic 
approach inevitable \cite{Leggett87,Weiss12}. 

Two questions of general current interest which can only be addressed based on microscopic modeling 
are: (1) How does the dissipative nonequilibrium dynamics of a quantum system with only a few degrees 
of freedom compare to the standard example of classical dissipation, the damped harmonic 
oscillator (DHO) and do such systems show a coherent-to-incoherent transition of the same type as 
it is found in the classical case? Studying the time-evolution of the ohmic spin-boson 
model (SBM) \cite{Leggett87,Weiss12} we show analytically that the dynamics in the 
coherent regime as  well as the transition to the incoherent one are significantly different 
from their classical counterparts. 
(2) What is the role of non-markovian terms in the dynamics 
of quantum dissipative systems? By considering 
parameter quenches across the coherent-to-incoherent transition in the SBM we show 
that non-markovian memory of the state before the quench heavily affects the time 
evolution after it. When quenching from the incoherent to the coherent regime this 
effect can be so strong that the dynamics after the quench is monotonic; the coherent
oscillatory behavior is fully suppressed. We provide a qualitative analytical explanation of this 
numerical finding. For quenches in the opposite direction nonmonotonic behavior can be transferred 
deep into the incoherent part of the dynamics. 

{\it Model}---The SBM is arguably the most important model used 
to describe dissipation in small quantum systems coupled to an environment. 
Its Hamiltonian reads
\begin{eqnarray}
H = \frac{\epsilon}{2} \sigma_z - \frac{\Delta}{2} \sigma_x + \sum_k \omega_k b_k^\dag b_k^{} - \sum_k
 \frac{\lambda_k}{2} \sigma_z \left(  b_k^\dag + b_k^{}\right),  
\end{eqnarray}
with the Pauli matrices $\sigma_\nu$, $\nu=x,z$ and bosonic ladder operators $b_k^{(\dag)}$. 
A spin-1/2 with Zeeman splitting $\epsilon$ and tunneling $\Delta \geq 0$
between the two states is coupled by $\lambda_k$ to a reservoir of bosonic 
modes with dispersion $\omega_k$. The spin-boson coupling is characterized 
by a spectral density $J(\omega) = \sum_k \lambda_k^2 \delta(\omega-\omega_k)$ given 
by the details of the model underlying the SBM \cite{Leggett87}. 
The dynamics was mainly studied in the ohmic case with $J(\omega) = 2 \alpha \omega 
\Theta(\omega_c - \omega)$ and coupling $\alpha\geq 0$ 
\cite{Leggett87,Weiss12,Egger94,Egger97,Lesage98,Anders06,Wang08,Orth10A}.
To investigate the fundamental questions raised above we reduce the number of parameters 
by considering the 
case $\epsilon=0$, the scaling limit, with the 
high-energy cutoff $\omega_c$ being much larger than any other scale, and temperature $T=0$.
In this strong-coupling limit the many-body physics becomes most intriguing and  
perturbative approaches fail \cite{Leggett87,Weiss12}. 
Other parameter regimes are investigated in 
Refs.~\cite{Grifoni98,Keil01,DiVincenzo05,Hackl08,Alvermann09,Orth10B,Orth12,Kast13a,Kast13b}. 
It is established that the time evolution 
of the spin expectation value \cite{footnoteTLS} $P(t)=\left< \sigma_z(t) \right>$ changes from being 
coherent, that is damped oscillatory,
for $\alpha<1/2$ to incoherent, that is monotonically decaying, for 
$1/2 < \alpha < 1$.

In studies of the nonequilibrium dynamics the system is often assumed 
to be in an initial product state of spin-up and the boson 
vacuum. The time evolution is performed for $\alpha>0$ and $P(t)$ 
asymptotically tends to zero. In the following we refer to this setup as the 
relaxation protocol. We furthermore consider two quench protocols. 
In the first one the system is prepared in the above initial state 
and the time evolution is performed with a coupling $2\alpha_i - 1> 0$ 
(incoherent regime) up to $t=t_q$. At $t_q$ it is abruptly switched 
to $\alpha_f$, with $2\alpha_f -1 = 1 - 2\alpha_i < 0$ (coherent regime), 
and the evolution is continued. In the second 
quench protocol we proceed in opposite order and quench from $2\alpha_i - 1 < 0$ to 
$2\alpha_f -1 = 1 - 2\alpha_i >0$. 

{\it Relaxation protocol}---The dynamics of the SBM 
for couplings sufficiently away from the transition at $\alpha=1/2$ 
is understood in detail. 
For small $\alpha$, that is deep in the coherent regime, it compares well to the 
one of the classical DHO \cite{Anders06,Wang08,Orth10A,Orth12} 
which is given as the sum of terms each being the product of an exponentially 
damped and an oscillatory factor. Deep in the incoherent regime 
the dynamics of the SBM is dominated by a single exponentially decaying 
term \cite{Anders06,Wang08} in resemblance to that of the DHO. 

\begin{figure}[t]
\includegraphics[width=0.95\linewidth,clip]{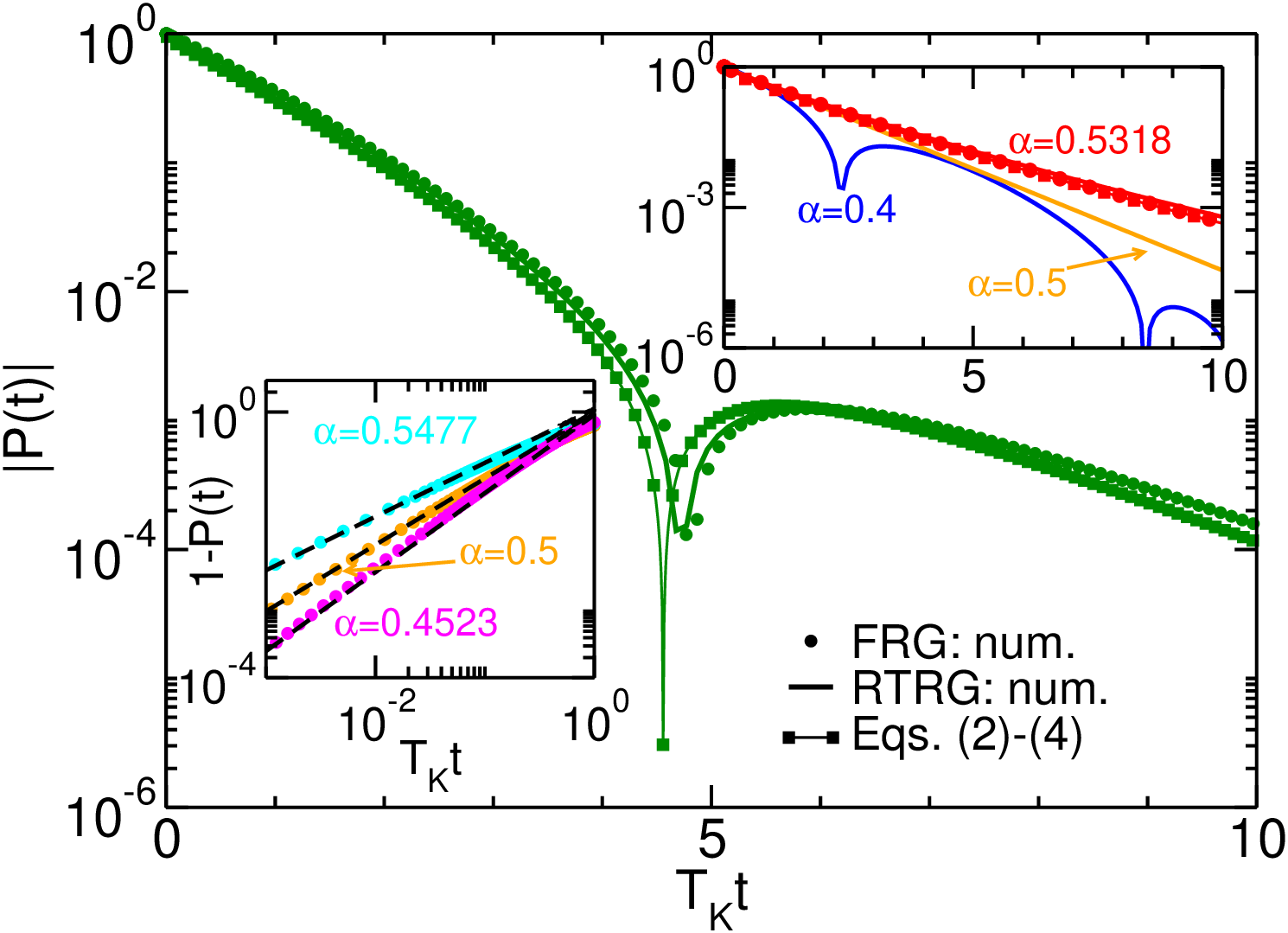}\vspace*{-0.2cm}
\caption{(Color online) Time evolution of a spin-1/2 
prepared in its up state and coupled to an ohmic boson bath at $t=0$. 
The spin  expectation value $|P(t)|$ is shown for different couplings $\alpha$. 
Dips correspond to zeros; to make those visible we use a log  $y$-axis scale. 
Main panel: comparison of the numerical solutions 
of the FRG and RTRG equations as well as the analytical result Eqs.~(\ref{branch0})-(\ref{branchpole}) 
for $\alpha=0.4682$. Right inset: the same as in the main panel but for various $\alpha$ 
(analytical expression and FRG for $\alpha=0.5318$ only; for this coupling the curves are barely 
distinguishable). The dynamics in the coherent regime 
is very different to that of the classical damped oscillator. Left inset: FRG data for $1-P(t)$ at short 
times (dots) compared to the NIBA prediction (dashed lines).}
\label{fig1}
\end{figure}

We now address if the analogy also holds close to the coherent-to-incoherent 
transition. For a controlled access to this 
intruiging regime we employ the smallness of $g=1-2 \alpha$. 
The problem is characterized by the low-energy scale 
$T_K =\Delta\left(\Delta / \omega_c\right)^{\alpha/(1-\alpha)}$ \cite{Leggett87,Weiss12},
which we take as our unit of energy and measure time in units of $T_K^{-1}$. 
Our analytical result for $P(t)= P_{\rm bc}(t) + P_{\rm p}(t)$ at $|g| \ll 1$ and $t \gtrapprox 1$ 
reads 
\begin{align}
\mbox{} \hspace{-.2cm} P_{\rm bc} (t)  & \approx \frac{1}{\pi} 
\mbox{Im}\Bigl\{ e^{-\gamma t} \mbox{E}_1\Bigl(\left[\tfrac{1}{2}\Gamma_{2}^\ast-\gamma\right] t\Bigr) 
\Bigr\} , \;\; \gamma = e^{- i \pi g} t^g ,
\label{branch0} \\
\mbox{} \hspace{-.2cm} P_{\rm p} (t)   & \approx    2 \frac{1-g}{1+g}  \cos \left( \Omega t \right) e^{-\Gamma_1^\ast t} \Theta(g) ,
\label{pole0} \\
\Gamma_2^\ast & \approx 2 \left[\frac{\pi g}{2 \sin (\pi g)} \right]^{\frac{1}{1+g}}, \;\;
\Omega+i\Gamma_1^\ast \approx e^{(i\pi+\ln 2)\frac{g}{1+g}} ,
\label{branchpole}
\end{align}
with the exponential integral $E_1$. It constitutes one of the main results of this Letter. 
As outlined below it is obtained employing a two-step procedure which is based on 
complementary renormalization group (RG) approaches: the functional (F) \cite{Metzner12} and the 
real-time (RT) \cite{Schoeller09} RG. 
In vast contrast to the dynamics of the classical DHO 
which on the coherent side shows infinitely many zeros even very 
close to the coherent-to-incoherent transition $P(t)$ only features a 
single zero for $\alpha$ close to $1/2$. This is exemplified in the main panel of 
Fig.~\ref{fig1} which shows $|P(t)|$ for $\alpha = 0.4682$; 
no additional zeros are found for times larger than 
the ones shown. Our analytical result is compared to the numerical solutions 
of the differential flow equations (see below) of the FRG 
and the RTRG. 
The excellent agreement of all three curves indicates that we fully 
control our approximations.  The right inset shows $|P(t)|$ for different $\alpha$. 
In the incoherent regime the three curves coincide even better 
($\alpha = 0.5318$).  
For $\alpha$ deeper in the coherent regime further zeros appear, e.g.~in total two for 
$\alpha=0.4$. 

A `pole contribution' $P_{\rm p}$ Eq.~(\ref{pole0}) to $P$ 
is also found in two alternative analytical approaches: the noninteracting 
blip approximation (NIBA) \cite{Leggett87,Weiss12} and conformal field 
theory (CFT) \cite{Lesage98}. Our ratio $\Omega/\Gamma_1^\ast$ 
Eq.~(\ref{branchpole}) coincides with the one 
obtained by these methods \cite{footnote4}. While no 
 `branch-cut contribution' $P_{\rm bc}$ Eq.~(\ref{branch0}) appears in 
CFT \cite{Lesage98} the one of NIBA leads to a purely algebraic decay which 
turned out to be an artifact of this approximation \cite{Weiss12}. In Ref.~\onlinecite{Egger97} 
NIBA was improved leading to 
\begin{eqnarray}
P_{\rm bc} (t) = - g [1+ 3 \Theta(-g)] \frac{e^{-t/2}}{t^{1+|g|}} . 
\label{bcimprovedNIBA}
\end{eqnarray}
This result agrees to our Eq.~(\ref{branch0}) evaluated for $t^{|g|} \gg 1$ and 
taking into account that to leading order $\Gamma_2^\ast  \approx 1$ (that is $T_K$). 
We emphasize that for $\alpha < 1/2$ the sum of the pole contribution 
Eq.~(\ref{pole0}) and the asymptotic result Eq.~(\ref{bcimprovedNIBA}) does not give 
a meaningful approximation to the dynamics for the times shown in Fig.~\ref{fig1} for 
which the coherent and incoherent parts are comparable
and nonmonotonic (`oscillatory') behavior is found. 
For those it is inevitable to keep the branch-cut term in the form of Eq.~(\ref{branch0}). 
Thus none of the existing analytical 
approaches to the SBM allows to uncover the crucial difference between the dissipative 
dynamics of the classical DHO and the SBM close to the coherent-to-incoherent 
transition. Furthermore, none of the numerical methods applied to the 
SBM \cite{Anders06,Wang08,Orth10A,Orth12} was used to investigate 
the dynamics in the transition region. We speculate that the data 
are to noisy to address questions of the above type;
note that the $y$-axis of Fig.~\ref{fig1} covers 
six orders of magnitude.  

Further confidence in our RG methods can be gained from considering
the short time dynamics $t \lessapprox 1$. For this NIBA predicts \cite{Weiss12}
$1- P(t) =  t^{1+g}/\Gamma(2+g) +{\mathcal O}(t^{2+2g})$ which favorably compares 
to our numerical results (left inset of Fig.~\ref{fig1}). In fact, 
within both our RG approaches this expression can be derived analytically as well.  

{\it Methods}---We next briefly describe our methods. Readers interested 
in results only can skip this part. 

Using our RG approaches we do not 
directly study the SBM but employ the mapping to the fermionic 
interacting resonant level model (IRLM) \cite{Leggett87,Weiss12}. It is sufficient 
to know that the coupling $\alpha$ of the SBM is related to the two-particle 
interaction $U$ of the IRLM: $1-2\alpha = 2U - U^2=g$. 
The case $\alpha=1/2$ corresponds to the noninteracting resonant level model ($U=0$) 
and is exactly solvable. The FRG flow equations for the Keldysh components of the 
many-body self-energy of the IRLM derived in Refs.~\onlinecite{Kennes12} 
(relaxation protocol) and \onlinecite{Kennes12a} (quench protocols) are 
controlled to leading order in $U$, that is $g$, and can directly be applied 
to the SBM.  

\begin{figure}[t]
\includegraphics[width=0.95\linewidth,clip]{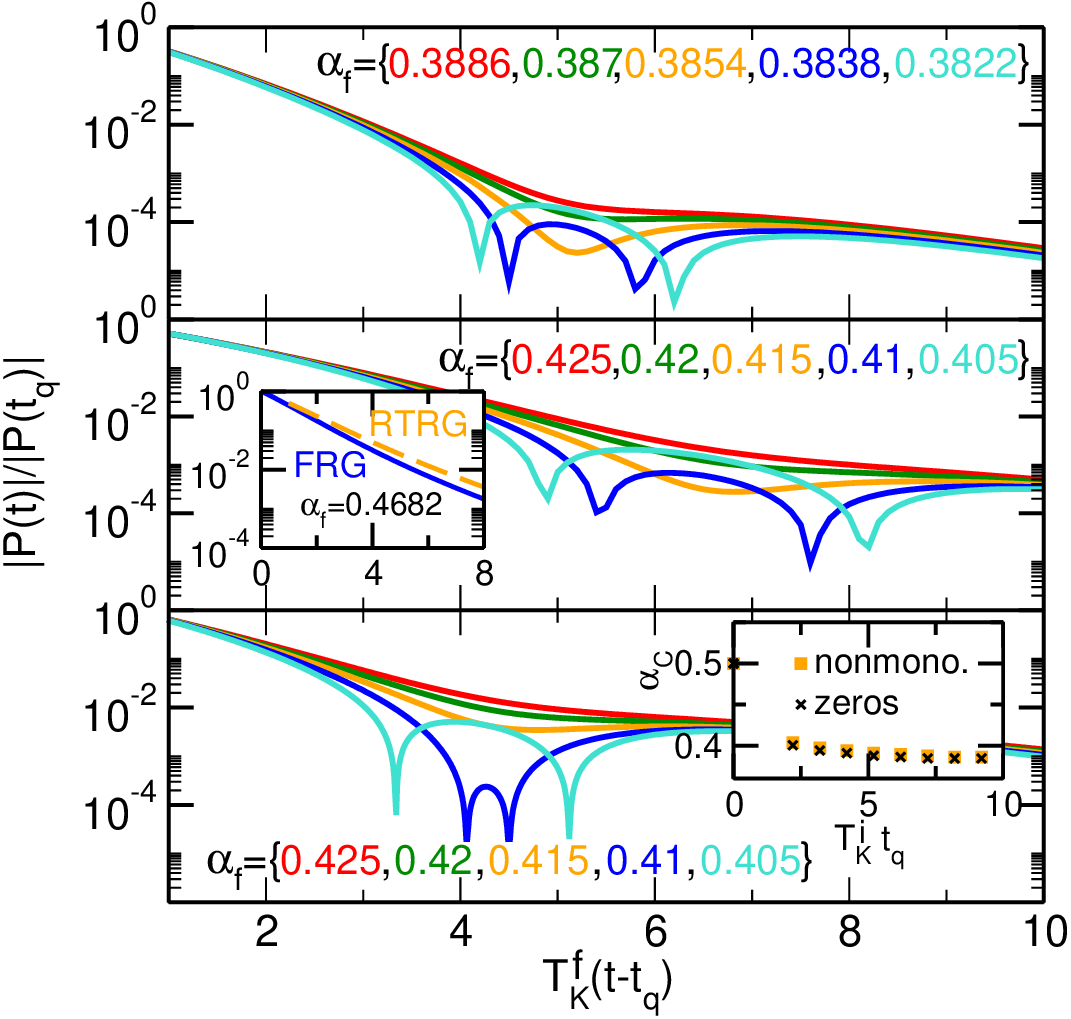}\vspace*{-0.2cm}
\caption{(Color online)  Spin expectation value $P(t)$ of a spin-1/2 with 
the initial time evolution given by the spin-boson Hamiltonian and a coupling
from  the incoherent regime $\alpha_i>1/2$. At $t=t_q$ it is switched to 
$\alpha_f<1/2$ from the coherent one. 
The scaling limit is realized by $\Delta/\omega_c=1/200$. 
Times are restricted to $t>t_q$; for $0 < t < t_q$, see Fig.~\ref{fig1}.
Left inset: comparison of numerical FRG and RTRG data; the non-markovian memory 
completely suppresses the coherent `oscillatory' behavior. It only survives for 
$\alpha_f$ smaller than a critical coupling $\alpha_c$; see the   
upper [numerical solution of FRG equations], central [numerical solution of RTRG equations], and 
lower [analytical result Eq.~(\ref{anaquench})] panels.  
Right inset: dependence of $\alpha_c$ on the time $T_K^i t_q$ the evolution was 
performed in the incoherent regime (FRG data). Two different definitions of $\alpha_c$ 
are compared. In the first $\alpha_c$ is defined as the $\alpha_f$ at which the 
first zero of $P(t)$ can be observed, in the second as the $\alpha_f$ at which the 
curve first shows a nonmonotonicity. The two values barely differ.}
\label{fig2}
\end{figure}

\begin{figure}[t]
\includegraphics[width=0.95\linewidth,clip]{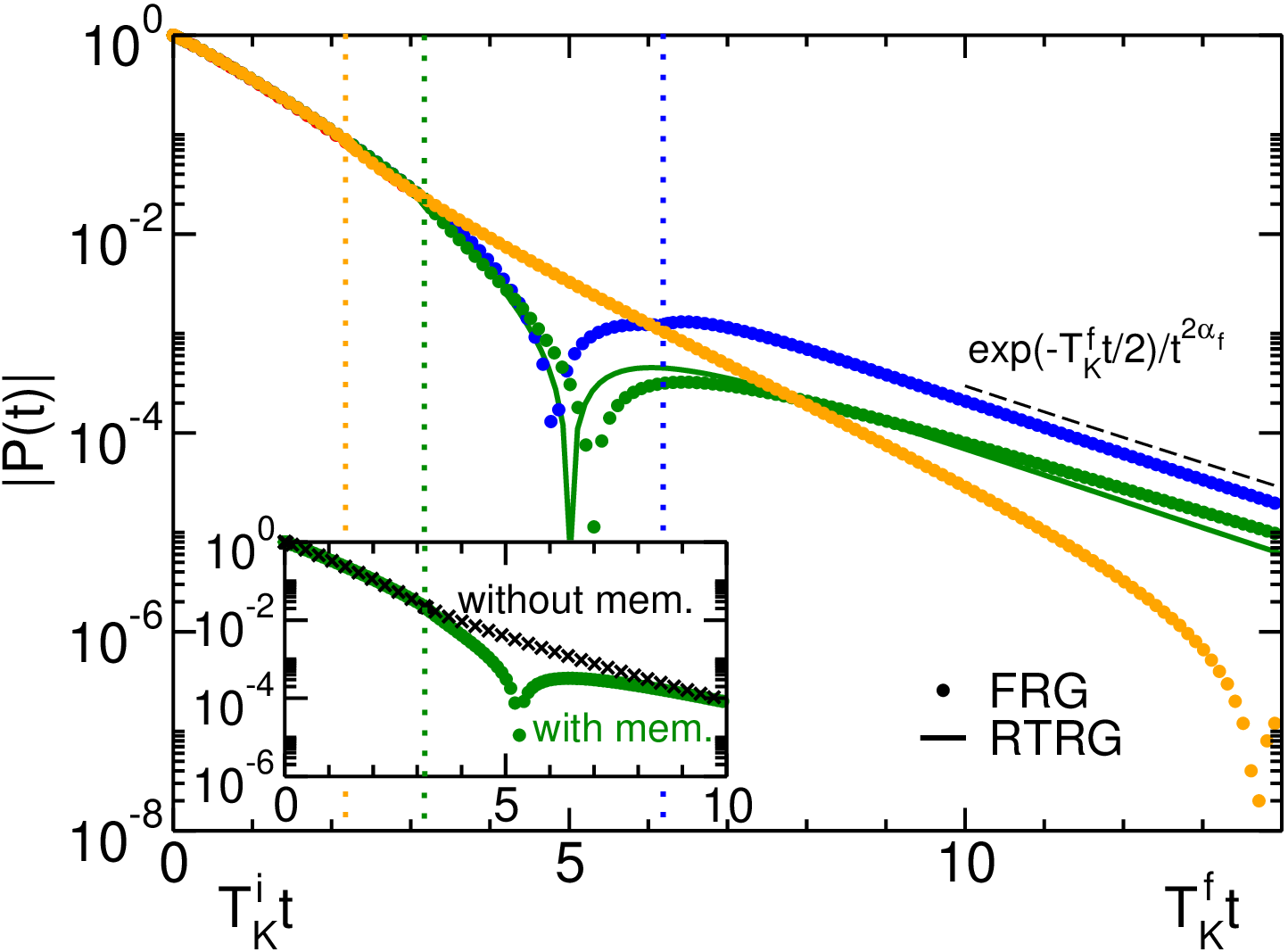}\vspace*{-0.2cm}
\caption{(Color online) The same as in Fig.~\ref{fig2} but for quenches 
from the coherent to the incoherent regime. The spin expectation value 
$|P(t)|$ for $\alpha_i=0.4682$ and different quench times $t_q$ (indicated by 
the vertical dotted lines) is shown. At the respective $t_q$ the $x$-axis 
scale is switched from $T_K^it$ to $T_K^f t$. Main panel: data of the  
numerical solutions of the FRG and RTRG flow equations. The 
thin dashed line is an exponential term with rate $T_K^f/2$ and a 
subleading power-law correction. 
Inset: FRG data with and without the memory term.}
\label{fig3}
\end{figure}

In the complementary RTRG \cite{Schoeller09} one focuses on the reduced 
density matrix of the local system and describes its dynamics in 
Liouville-Laplace space by
\begin{eqnarray}
P(t) = \frac{i}{2 \pi} \int_{-\infty+i0^+}^{\infty+i0^+} dE \,e^{-iEt}\, \Pi_{1}(E) ,
\label{integral}
\end{eqnarray}  
where $\Pi_1(E)=[E+i\Gamma_1(E)]^{-1}$ is an effective propagator; in the IRLM $\Gamma_1(E)$ denotes the
charge relaxation rate. The Laplace variable $E$ can be used as the flow
parameter \cite{Pletyukhov12}, leading to 
\begin{equation}
{d\Gamma_{1/2}(E)\over dE}\,=\,-g\,\Gamma_1(E)\,\Pi_{2/1}(E),
\label{flowequ}
\end{equation} 
with $\Pi_2(E)=[E+i\Gamma_2(E)/2]^{-1}$ and initial values 
$\Gamma_{n}(i \omega_c) =\Delta^2/\omega_c $; in the IRLM $\Gamma_2/2$ describes the level broadening.
Solving Eq.~(\ref{flowequ}) offers the unique possibility to 
identify the singularities of the propagator (poles and branch cuts) in the 
lower half of the complex plane, from which the individual terms of the time 
evolution can be analyzed. The RG equations (\ref{flowequ}) contain all terms 
${\mathcal O} \left(U{\Delta^2\over \omega_c E}\right)$. However, some terms 
${\mathcal O}\left(U \left[{\Delta^2\over \omega_c E}\right]^2\right)$ are neglected. This must 
be contrasted to our FRG approach, which contains all orders in $\Delta^2/\omega_c$. To verify 
that for small $|U|$, that is small $|g|$, considering Eq.~(\ref{flowequ}) is sufficient,
we always compare the numerical solutions of the FRG and RTRG flow equations 
(see Figs.~\ref{fig1}-\ref{fig3}).

In an analytical solution of our RTRG equations (\ref{flowequ}) one has to separately 
consider the branch-cut contribution to the integral Eq.~(\ref{integral}) for all $g$ 
and the pole contributions for $g \geq 0$. 
The former follows from the  branch cut of $\Gamma_1(E)$ on the imaginary 
axis starting at the pole $-i \Gamma_2^\ast/2$ of $\Pi_2(E)$. Taking 
$\Gamma_2(E)\approx \Gamma_2^\ast$ in $\Pi_2(E)$, we get 
$\Gamma_1(E) \approx (\Gamma_2^\ast/2 - i E )^{-g}$ from Eq.~(\ref{flowequ}), leading 
to 
\begin{eqnarray}
P_{\rm bc} (t) = - \frac{e^{-\Gamma_2^\ast t/2}}{\pi} \int_0^\infty  \!\! dx \,
\mathrm{Im} \left\{ \frac{e^{-xt}}{\frac{\Gamma_2^\ast}{2} + x - e^{i \pi g} x^{-g}} \right\} .
\label{branchint}
\end{eqnarray}
Replacing $x^g$ by $t^{-g}$ in the denominator gives a very good approximation for 
$t \gtrapprox 1$ and leads to Eq.~(\ref{branch0}).
A comparison with the numerical solution shows that to find the 
pole positions $z_1=\pm\Omega-i\Gamma_1^\ast$ [$z_2=-i\Gamma_2^\ast/2$] of
$\Pi_{1}(E)$ [$\Pi_{2}(E)$] one can set $\Gamma_1(E)\approx\Gamma_2(E)\approx iz_1$ 
[$\Gamma_1(E)\approx\Gamma_2(E)\approx\Gamma_2^\ast$] 
in $\Pi_n(E)$. The RG equations can then be solved analytically 
leading to Eq.~(\ref{branchpole}) as well as the pole contribution to $P(t)$ 
Eq.~(\ref{pole0}).

{\it Quench protocol 1}---We next investigate the role of non-markovian memory in 
dissipative dynamics. In Fig.~\ref{fig2} we show $P(t)$ when 
quenching at $t_q$ from the incoherent to the coherent regime.  
The two different couplings $\alpha_i$ and $\alpha_f$ lead to the two characteristic 
scales $T_K^i$ and $T_K^f$, which in the scaling limit differ by orders of magnitude.
Therefore the model parameters $\Delta$ and $\omega_c$ cannot be 
scaled out as efficiently as in the relaxation protocol (by taking 
$T_K$ as the unit of energy). 
To minimize the influence of the transient dynamics before the quench we consider 
$T_K^i t_q$ between 5 and 10.
In the left inset we compare the data obtained by the numerical solution of the 
FRG \cite{Kennes12a} and  RTRG equations (see below). The agreement 
for $\alpha_{i/f}$ close to $1/2$ is excellent. Even though the same coupling 
$\alpha_f = 0.4682$ as in the main panel of Fig.~\ref{fig1} is considered the data 
for the quench dynamics displayed in the left inset of Fig.~\ref{fig2} are monotonic; no 
indication of coherent behavior is found. The incoherent dynamics before the quench thus 
heavily affects the one afterwards. 
To further investigate this, results for smaller $\alpha_f$ are shown in the upper 
(FRG) and central (RTRG) panel.  
Nonmonotonic (`oscillatory') behavior can only be found for $\alpha_f$ being smaller than some 
critical coupling $\alpha_c$. As $|1-2\alpha_c| \ll 1$ does not hold strictly, our two 
approximate RG methods give values for $\alpha_c$ which differ by a few percent. 
Its precise value can only be obtained using a method which treats higher order 
contributions consistently. 

Quenches were so far not studied using RTRG; the technical 
details will be given elsewhere. In this approach the memory 
of the spin state is preserved in the excitations of the bath. 
Classifying the memory contributions by the number of bath 
excitations that have gone through the quench
one can show that the terms with $n$ excitations are 
proportional to $A^{n}=(T^i_K/T^f_K)^{n/2} \approx (\Delta/\omega_c)^{2n|g_f|}$, with $A \ll 1$. 
Restricting ourselves to the case of a single excitation with Matsubara frequency $\Lambda$ we 
obtain ($t'= t -t_q \geq 0$)
\begin{eqnarray}
\mbox{} \hspace{-.3cm} P(t') \!\!  & = & \!\! P^f(t') P^i(t_q) - \!  
g_i \! \! \int_0^\infty \!\! d \Lambda F_\Lambda^f(t')  F_\Lambda^i(t_q) , \label{nonlocal} \\ 
\mbox{} \hspace{-.3cm} F_\Lambda^\kappa(t) \!\!  & = &  \!\! - \int_{- \infty}^\infty \frac{dE}{2 \pi} 
e^{-iEt} \Pi_{1}^\kappa(E+i\Lambda)\sqrt{\Gamma_1^\kappa(E)}\Pi_{2}^\kappa(E) \nonumber
\end{eqnarray}
with $\kappa = i,f$ and $P^\kappa$, $\Gamma_{1/2}^\kappa$ computed 
as in the relaxation 
protocol with the corresponding $\alpha_{i/f}$ and initial condition 
$P^\kappa(0)=1$. The first term describes standard relaxation 
while the second memory term implies (dissipative)  
non-markovian dynamics. It has no analog for quenches in closed systems \cite{Polkovnikov11}.
The RTRG data of 
Fig.~\ref{fig2} were obtained by numerically performing the $E$ and $\Lambda$ 
integrals in Eq.~(\ref{nonlocal}) on the basis of the numerical 
solution of Eq.~(\ref{flowequ}). 
Fixing $\omega_c$  the results depend only very weakly on $\Delta^2/\omega_c$ (via $A$). Varying 
$(\Delta/\omega_c)^2$ by an order of magnitude around the average 
value $2.5\cdot 10^{-5}$ results in a change of $\alpha_c$ by only a few percent.

To gain qualitative analytical insight of the interplay of the relaxation dynamics and 
the non-markovian correction we evaluate the integrals keeping only the 
terms which dominate for $|g_{i/f}| \ll 1$. The result
\begin{eqnarray} 
\frac{P(t')}{P(t_q)} \approx  2 (1 -  A) e^{-\Gamma^{\ast f}_{1} t' }  \cos( \Omega^f t')   
  +  A e^{-\Gamma^{\ast f}_{2} t'/2}  
\label{anaquench}
\end{eqnarray}
is shown in the lower panel of Fig.~\ref{fig2}. The memory generates a 
coherent and an incoherent contribution $\propto A$. The first has a negative sign 
and suppresses the coherent part while the second enhances the incoherent term 
compared to the one of the relaxation protocol; for $g_f \ll 1$ the latter is 
subdominant and thus not written in Eq.~(\ref{anaquench}). This explains the
appearance of a critical $\alpha_c$.

{\it Quench protocol 2}---We finally discuss the dynamics when 
quenching at $t_q$ in the opposite direction, that is from the coherent to the 
incoherent regime.  To keep the discussion transparent we focus on 
$\alpha_i$'s  for which only a single zero at time $t_0$ is found 
in the relaxation protocol.  In Fig.~\ref{fig3} we show $|P(t)|$ obtained from the numerical 
solution of FRG and RTRG equations for $\alpha=0.4682$. In the relaxation 
protocol $T_K t_0 \approx 5$ for this coupling (see Fig.~\ref{fig1}).  
Different quenching times $t_q$ are considered. For $t_q > t_0$ the 
dynamics at times larger than $t_q$ very quickly adapts to the new 
rate $\approx T_K^f /2$ of the incoherent dynamics [see Eq.~(\ref{bcimprovedNIBA}) 
and the dashed line in Fig.~\ref{fig3}]. 
The behavior is significantly different for $t_q < t_0$. In this case nonmonotonic (`oscillatory') 
behavior is found for times $t>t_q$ at which the time evolution is performed with 
$\alpha_f=0.5318>1/2$. The smaller $t_q - t_0$  the further the zero is transfered into the 
incoherent regime; similarly to the first quench protocol the memory persist over 
several $1/T_k^f$. In the inset of Fig.~\ref{fig3} it is shown that the zero of $P(t)$ 
vanishes if the memory term is switched off. This indicates that the appearance of the zero is 
exclusively coded in the history of the dynamics and not in the values of the systems density 
matrix at a selected time. 

{\it Summary}---We have shown that the coherent dynamics of the ohmic SBM, being 
the prototype model of dissipative quantum mechanics, close to the coherent-to-incoherent 
transition is very different from the one of the classical DHO. Our study
furthermore revealed the crucial importance of non-markovian memory in the nonequilibrium 
time evolution when quenching across this transition. 

\textit{Acknowledgments}---This work was supported by the DFG via FOR 723. We thank 
R.\ Egger, C. Karrasch, and U.\ Weiss for discussions.

{}


\begin{thebibliography}{}
\bibitem{footnoteTLS} $P$ is equivalent to the population imbalance when viewing
the SBM as a two-level system \cite{Leggett87,Weiss12}.
\bibitem{footnote4}
The prefactor of $P_{\rm p}(t)$  Eq.~(\ref{pole0}) agrees with the one of NIBA only to order $g^0$. We were informed 
by U.\ Weiss that computing it within improved NIBA \cite{Egger97} gives a result which agrees 
to ours to order $g$.
\end{thebibliography}
\end{document}